\documentclass[a4paper,11pt]{JHEP}
\usepackage[T1]{fontenc}
\usepackage{ae,aecompl,aeguill}	
\usepackage[ansinew]{inputenc}
\usepackage{amsmath}
\usepackage{amssymb}
\usepackage{slashed}
\usepackage{graphicx}
\usepackage{subfigure}

\newcommand{\be}{\begin{equation}}
\newcommand{\ee}{\end{equation}}
\newcommand{\beqa}{\begin{eqnarray}}
\newcommand{\eeqa}{\end{eqnarray}}

\newcommand\CLASS{{\tt CLASS}~}
\newcommand\CAMB{{\tt CAMB}~}
\newcommand\CMBFAST{{\tt CMBFAST}~}
\newcommand\CMBEASY{{\tt CMBEASY}~}
\newcommand\RECFAST{{\tt RECFAST}~}

\preprint{CERN-PH-TH/2011-083, LAPTH-011/11} \title{The Cosmic Linear
Anisotropy Solving System\\ (CLASS) III: Comparision with CAMB for
$\Lambda$CDM} \author{Julien Lesgourgues$^{a,b,c}$\vspace{.2cm}\\
{$^a$}Institut de Th\'eorie des Ph\'enom\`enes Physiques,\\ \'Ecole
Polytechnique F\'ed\'erale de Lausanne,\\ CH-1015, Lausanne,
Switzerland.\vspace{.2cm}\\ {$^b$} CERN, Theory Division,\\ CH-1211
Geneva 23, Switzerland.\vspace{.2cm}\\ {$^c$} LAPTh (CNRS -
Universit\'e de Savoie), BP 110,\\ F-74941 Annecy-le-Vieux Cedex,
France.}  

\abstract{By confronting the two independent Boltzmann codes \CLASS
  and {\tt CAMB}, we establish that for concordance cosmology and for
  a given recombination history, lensed CMB and matter power spectra
  can be computed by current codes with an accuracy of 0.01\%. We list
  a few tiny changes in \CAMB which are necessary in order to reach
  such a level. Using the common limit of the two codes as a set of
  reference spectra, we derive precision settings corresponding to
  fixed levels of error in the computation of a CMB likelihood. We
  find that for a given precision level, \CLASS is about 2.5 times
  faster than \CAMB for computing the lensed CMB spectra of a
  $\Lambda$CDM model. The nature of the main improvements in CLASS
  (which may each contribute to these performances) is discussed in
  companion papers.} %\keywords{}

\begin{document} 

\section{Introduction}

In a few companion papers \cite{class_general,class_approx}, we expose
the main motivations for developing and releasing a new Boltzmann
code, the Cosmic Linear Anisotropy Solving System ({\tt
CLASS})\footnote{available at {\tt http://class-code.net}}. Our
targets are {\it (i)} friendliness and flexibility, {\it (ii)} a full
control of errors, {\it (iii)} computational speed, {\it (iv)} to get
an opportunity to test the accuracy of other codes with an independent
one. In this paper, we will focus on the last three aspects. Since
\CAMB \cite{Lewis:1999bs} is the only public code which has been
regularly maintained over the past few years (including important
updates on the recombination side), and since it is completely
independent from \CLASS (see however the disclaimer below about
recombination), we only compare here \CLASS with \CAMB (last available
version of January 2011). For a recent comparison between \CAMB and
\CMBFAST \cite{Seljak:1996is}, which are {\it not} completely
independent, we refer the reader to
ref.~\cite{Hamann:2009yy}. Precision tests for \CMBEASY were presented
in \cite{Doran:2003sy}, and for other codes in \cite{Seljak:2003th};
however, these last two papers are now out-dated: since 2003,
significant progresses were made on the side of recombination, lensing
calculation, etc.; besides, in these previous papers, the accuracy of
the codes was pushed to a much smaller level than in
\cite{Hamann:2009yy} and in the present work. The authors of
\cite{Seljak:2003th} showed that the {\it unlensed} CMB spectra of
several codes agree almost to the 0.1\% level. This observation was
crucial for interpreting WMAP data, but for Planck and post-Planck
data (CMB, cosmic shear surveys, etc.) we would like to push the
comparison further, in order to derive reference settings in which the
theoretical error is proved to be much smaller than the observational error.

Hence, the first goal in this paper is to check the agreement between
\CLASS and \CAMB, and to identify possible systematic errors
induced by one or both of them. This investigation is presented in
section \ref{agreement}.  The discrepancy which will remain after
tracking all possible sources of inaccuracy will give an estimate of
the absolute precision of current Boltzmann codes. Up to this level of
accuracy, the spectra produced by any of the codes can be treated as
reference spectra. The most robust way to estimate the accuracy of
\CLASS or \CAMB for a given set of precision parameters is to compare
the output spectra with these reference spectra. In section
\ref{precision}, we will use this method for establishing a few sets
of well-calibrated precision parameter sets. Finally, in section
\ref{performance}, we will compare the speed of the two codes for a fixed
level of accuracy.

For concision, we restrict this work to the consideration of minimal
$\Lambda$CDM models, and to the calculation of the CMB spectra and
matter power spectrum $P(k)$ for scalar modes with adiabatic initial
conditions.  Within the small range of $\Lambda$CDM parameters allowed
by current cosmological data, there is no reason for which the
precision of the code would vary significantly with the cosmological
parameters: hence, we can base our entire comparison on just one set
of cosmological parameters close to the best-fit concordance
model. The \CLASS vs. \CAMB comparison for isocurvature modes, tensors or
extended cosmologies (e.g. with spatial curvature or massive
neutrinos) is left for future specialised communications (the case of
massive neutrinos and non-cold dark matter relics is already discussed
in a companion paper \cite{class_ncdm}).

In our estimate of the speed of the code, we will focus only on the
running time needed on one CPU for computing the CMB spectra in a
$\Lambda$CDM model. Any other comparison (including the matter
power spectrum, massive neutrinos, parallel runs, etc.) is likely to
be even more in the favor of {\tt CLASS}, thanks to its extended
approximation schemes dealing with simplified equations during
matter domination \cite{class_approx}, with fluid approximations for
non-cold relics \cite{class_ncdm}, with its advanced stiff integrator
\cite{class_approx}, with its large amount of parallelised loops, etc.

Of course, the physical equations integrated in \CLASS and \CAMB in
absence of any approximation scheme are the same, and \CLASS uses the
exact ``line-of-sight integral method'' proposed by Seljak \&
Zaldarriaga \cite{Seljak:1996is} and implemented in \CMBFAST, \CAMB and
\CMBEASY. \CLASS and \CAMB are still independent from each other in the
sense that everything in \CLASS apart from the recombination module has
been written from scratch in another language, using different overall
architecture, numerical algorithms, discretisation schemes,
approximations, etc. Apart from the exact physical equations, the only
common block is the module \RECFAST \cite{Seager:1999bc} (currently at
version 1.5 \cite{Scott:2009sz}) which computes the recombination
history for a given set of cosmological parameters. In {\tt CLASS}, the
Fortran version of \RECFAST v1.5 has been translated into C; it has
then been slightly modified by introducing smoothing functions in
order to avoid discontinuities in the derivatives of thermodynamical
variables (this turns out to be useful, see the related discussion
in section \ref{unlensed}); but apart from smoother transitions
between various approximation schemes, the equations of \RECFAST are
unchanged.

Simulating recombination is a tough problem which has its own level of
uncertainty. Experts agree that \RECFAST v.1.5 is accurate enough for
analysing Planck data, but it is clear that in this paper, we will
push the precision of the two Boltzmann codes at a level comparable or
even larger than the expected precision of {\tt RECFAST}. As long as \CLASS
and \CAMB use the same recombination code, it is possible to compare
them up to arbitrarily high level. But it should be clearly understood
that the goal of this paper is to estimate the accuracy of \CLASS and
\CAMB {\it for a given recombination history}. For concision, we will
not write such a disclaimer in all relevant places in the
text. Estimating the propagation of uncertainties from recombination
to the CMB spectra is an independent task, actually requiring
negligible errors from the Boltzmann code itself. In order to extract
cosmological parameters from highly accurate data (like that from
Planck or from future cosmic shear surveys), cosmologists need to be
sure that neither recombination nor Boltzmann codes can introduce
significant errors. The present work is only addressing
the Boltzmann side.

\section{Agreement with \CAMB\label{agreement}}

All along this work, we fix the $\Lambda$CDM cosmological parameters
to typical $\Lambda$CDM values\footnote{$h=0.7, T_{cmb}=2.726{\rm K},
\Omega_b=0.05, N_{eff}=3.04, \Omega_{cdm}=0.25, \Omega_{k}=0,
Y_{He}=0.25, z_*=10, A_s=2.3\times10^{-9} {\rm ~at~} 0.05~{\rm
Mpc}^{-1}, n_s=1$, exact same reionisation function $f(x_e)$ with a
``reionisation exponent'' (controlling the sharpness of reionisation)
equal to 1.5 and ``reionisation width'' equal to 0.5. Note that for an
accurate comparison, one must ensure that both \CAMB and \CLASS
write in output files with a sufficient number of digits.}, and vary
the precision parameters of \CAMB and {\tt CLASS}. We first need to
check that each code tends towards stable results when the precision
increases, and that the limits from the two codes agree with each
other. If this turns out to be the case, we will conclude that the
codes are not plagued by bugs or numerical issues, and we will be able
to interpret the obtained spectra as ``reference spectra''. Such a
conclusion is correct provided that the two codes do not contain the
same mistake, which would be very unlikely since they are independent
(in the sense discussed in the introduction).

  We start by tuning each \CLASS precision parameter one by one, up
  to a level at which the CMB spectra (unlensed/lensed temperature and
  E-polarisation, plus lensing potential spectrum) remain stable at
  the $0.01\%$ level in the range $2-3000$ (settings called {\it
    [CLASS:01]}). The corresponding precision parameter values can be
  found in the input file {\tt cl\_ref.pre} of the public distribution
  of the code (as explained in \cite{class_general}, \CLASS can be ran
  with two input files, one for cosmological parameters, and one for
  precision parameters: for instance, {\tt ./class my\_model.ini
    cl\_ref.pre}).

\subsection{Scalar unlensed CMB spectra\label{unlensed}}

Next, we run \CAMB with the following precision parameters (settings called
{\it [CAMB:01]}):

\vspace{0.2cm}

\begin{tabular}{rcl}
{\tt l\_max\_scalar}      &=& 3000\\
{\tt k\_eta\_max\_scalar}  &=& 12000\\
{\tt accurate\_polarization}   &=& T\\
{\tt accurate\_reionization}   &=& T\\
{\tt do\_late\_rad\_truncation}  &=& F\\
{\tt accuracy\_boost}          &=& 3\\
{\tt l\_accuracy\_boost}        &=& 3\\
{\tt l\_sample\_boost}          &=& 3
\end{tabular}

\vspace{0.2cm}

Note that setting all accuracy boost parameters to three is often used
in projects requiring high precision. Above, the parameter {\tt
k\_eta\_max\_scalar} is set to four times {\tt l\_max\_scalar}, while
the default recommendation is just two times. Hence, we expect these
settings to be very accurate.

We find nevertheless a small discrepancy in the unlensed temperature
multipoles $C_l^{TT}$'s, of the order of 0.7\% at small $l$, and 0.1\%
at large $l$ (see figure \ref{fig_cl}, left plot, curve labelled {\it
  CAMB:01/CLASS:01~T}). At this point, further increasing {\tt CAMB}'s
``accuracy boost'' parameters does not affect the results
anymore. Note that these parameters control {\it most} precision
parameters in {\tt CAMB}, but not {\it all} of them: several quantities are
fixed by hand inside each \CAMB module. We tried to identify these
quantities, to vary them one by one, and to check whether they have a
sizable impact. This turns out to be the case for at least two numbers
controlling:
\begin{enumerate}
\item the sampling of Bessel functions. \CAMB samples spherical Bessel
  functions $j_l(x)$ with four different linear steps in four
  intervals in $x$. For values $x>25$, $\Delta x$ is set to one. We
  reduced this value to 0.2 (settings called {\it [CAMB:02]}) and
  found that the $C_l$'s are affected by approximately 0.1\% (see
  figure \ref{fig_cl}, left plot, curve labelled {\it
    CAMB:02/CLASS:01~T}).
\item the sampling of the free electron fraction $x_e(\tau)$ along conformal time $\tau$, in view of computing
the visibility function $g(\tau)$. 
\end{enumerate}
This second issue deserves more explanations.  In {\tt CAMB},
thermodynamical quantities are computed in the following way: 
\begin{itemize}
\item[(a)] the \RECFAST module (version 1.5) computes the free electron
  fraction and baryons sound speed ($x_e(\tau)$, $c_b^2(\tau)$) by
  integrating the Boltzmann equations describing helium and hydrogen
  recombination over {\tt Nz} linear steps in redshift space, between
  $z=z_\text{max}$ and $z=0$. By default, $z_\text{max}$ = {\tt Nz} = 10'000,
  so the integration step is $\Delta z = 1$.
\item[(b)] the functions $x_e(\tau)$ is corrected to account for
reionisation (but the function $c_b^2(\tau)$ is not, as we shall see
in subsection \ref{pk}).
\item[(c)]
{\tt CAMB}'s ``{\tt ThermoData}'' module computes derived (and integrated) quantities like the
opacity $\kappa'$, the column density $e^{-\kappa}$, the visibility
function $g=\kappa' \exp[-\kappa]$ and its time-derivatives.
This last step is performed
using another linear sampling of these functions: the number of points
between $z_\text{max}$ and $0$ is then fixed by the integer {\tt nthermo}. By
default, {\tt nthermo} = {\tt Nz} = 10'000, so that the \RECFAST and
{\tt ThermoData} modules use the same sampling.
\end{itemize} In {\tt CLASS}, the step (a) is identical to {\tt CAMB}.
The step
(b) is also roughly similar, except for the
choice of $z$ sampling near the reionisation epoch (found
automatically for a given reionisation function), and the fact that
we ensure that $x_e(\tau)$ remains exactly continuous and
derivable at any time. The baryon sound speed is only computed after
this step.  Finally, for the (c) step, \CLASS always uses the $z$
sampling defined at steps (a) and (b).

By inspecting the thermodynamical quantities as a function of
redshift, we find that \CAMB and \CLASS agree very well as far as the
free electron fraction and opacity are concerned, while there are
small offsets in integrated and derived quantities like $e^{-\kappa}$
and $g$.  This signals some inaccuracy in the numerical integration,
derivation or interpolation routines used in at least one of the two
codes.

In order to check for the convergence of {\tt CAMB}'s thermodynamical
variables, we would like in principle to increase the values of {\tt
Nz} and {\tt nthermo}. In practise, increasing {\tt Nz} is not
possible due to the little discontinuities in the $x_e(\tau)$ and
$x_e'(\tau)$ functions returned by the raw version of \RECFAST v1.5: as
soon as the step decreases, spurious oscillations appear in all
derived quantities, and the final results diverge. However, it is
possible to keep {\tt Nz} fixed and to increase {\tt nthermo}, in
order to check that the integration/derivation of the function
$x_e(\tau)$ returned by \RECFAST is accurate enough. We find that
pushing {\tt nthermo} in the range 80'000-100'000 leads to stable
results, which are significantly different from those with the default
setting {\tt nthermo}=10'000.  Moreover, the new results are in
excellent agreement with the high-precision limit from {\tt CLASS}.
Note that increasing {\tt nthermo} but not {\tt Nz} has the
inconvenient effect of introducing little oscillations due to spline
interpolation, but they seem harmless for {\tt nthermo} $\sim$ 100'000,
since the spectrum does converge towards the one of {\tt CLASS}.

This suggest that in {\tt CAMB}, {\tt nthermo} should always be kept
of the order of 100'000 in order to gain one order of magnitude in
precision (from the 0.1\% to the 0.01\% level). At this step, there is
still a loophole in our reasoning. It might be the case that the two
codes agree by coincidence when {\tt nthermo}=100'000. Since we could
not push {\tt Nz} above 10'000 and {\tt nthermo} above 100'000, we
have not proved that the final result is fully converged with the
settings {\tt Nz} = 10'000, {\tt nthermo}=100'000.

Fortunately, this question can be addressed. Here comes into play the
fact that the \RECFAST implementation in \CLASS has a small difference
with respect to the one in \CAMB and the original one: it includes
smoothing functions in the system of differential equations, which
ensure that $x_e(\tau)$ is everywhere continuous and twice derivable.
These smoothing function do not affect $x_e(\tau)$ in a detectable
way. One can check it by changing the details (width and shape) of the
smoothing functions: this impacts the final results at a level much
below the 0.01\% level. Hence, the fact that \CLASS contains such
smoothing functions is not a possible explanation
for the small discrepancy discussed in the previous paragraphs.

If the impact of these smoothing functions is negligible, why did we
introduce them at all? The reason is purely numerical. \CLASS has a
unique parameter {\tt recfast\_Nz} which plays the role of both {\tt
Nz} and {\tt nthermo} in {\tt CAMB}. When we increase this parameter,
we increase the precision of the thermodynamics module without getting
spurious features, thanks to the smoothing functions. In other words,
the smoothing does not affect physical assumptions, but allows to test
convergence. When we increase {\tt recfast\_Nz} much above 100'000 in
{\tt CLASS}, the output spectrum remains very stable (it varies by
much less than 0.01\%). This is indeed a proof that \CLASS with {\tt
recfast\_Nz}=100'000, or \CAMB with {\tt Nz}=10'000 and {\tt
nthermo}=100'000 are both fully converged from the point of view of
thermodynamical quantities, at least at the 0.01\% level.

The residual difference between \CAMB with
{\tt nthermo} increased to 100'000 (setting {\it [CAMB:03]})
and \CLASS (settings {\it [CLASS:01]})
is shown in figure
\ref{fig_cl}, left plot, blue curve. The
difference is now sensitive to an additional increase in {\tt CAMB}'s
``accuracy boost'' parameters up to

\vspace{0.2cm}

\begin{tabular}{rcl}
{\tt accuracy\_boost}          &=& 12\\
{\tt l\_accuracy\_boost}        &=& 4\\
{\tt l\_sample\_boost}          &=& 3
\end{tabular}

\vspace{0.2cm}

\noindent (settings {\it [CAMB:04]}).  The results from {\it
  [CAMB:04]} and {\it [CLASS:01~T]} agree at a rather spectacular
  level of $0.01\%$ for $C_l^{TT}$ in the range $2-2000$. The only
  exception is the multipole $l=18$, for which the codes agree only at
  the $0.07\%$ level.  This detail is in any case completly irrelevant
  for practical purposes\footnote{the spike might be due to the value {\tt
  llmax} being fixed to 17 in {\tt CAMB}'s routine {\tt
  DoSourceIntegration}; we did not check this explicitely.}.  Note that
  {\tt CAMB}'s parameter {\tt do\_late\_rad\_truncation} can be
  indifferently set to T or F without affecting our conclusions: it
  governs an approximation which affects only the evolution of small
  scale perturbations after recombination, and do not impact
  significantly CMB observables.

For the same settings, the agreement between the polarisation spectra
$C_l^{EE}$'s is roughly as good as for temperature: the spectra differ
at most by $0.02\%$ in the range $40-3000$, or $0.04\%$ in the range
$3-40$ ($0.14\%$ for $l=2$).  The comparison between {\it [CAMB:04]}
and {\it [CLASS:01]} results for TT and EE is shown on the right
plot in Figure \ref{fig_cl}.

\FIGURE{
\includegraphics[width=7cm]{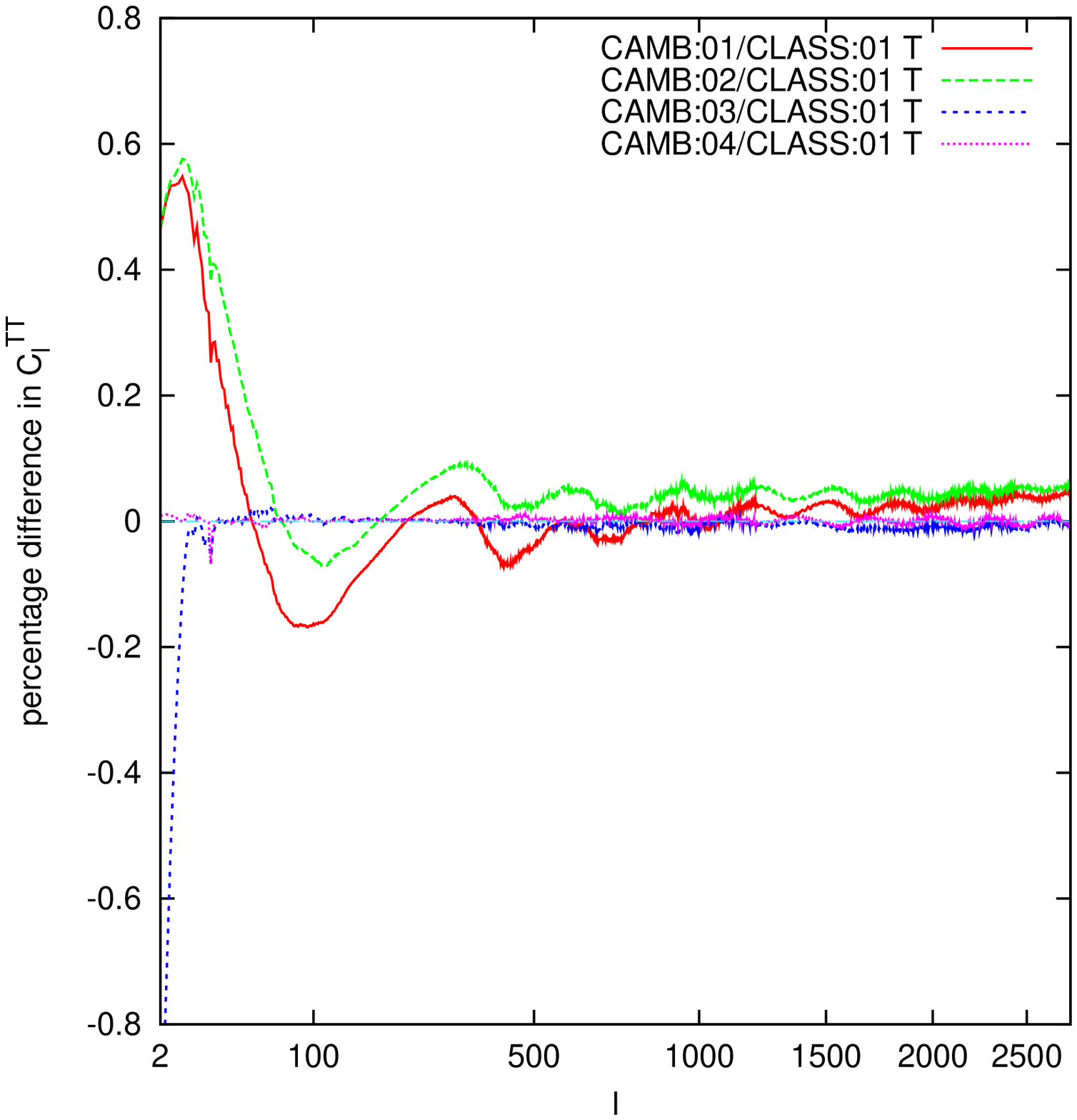}
\includegraphics[width=7cm]{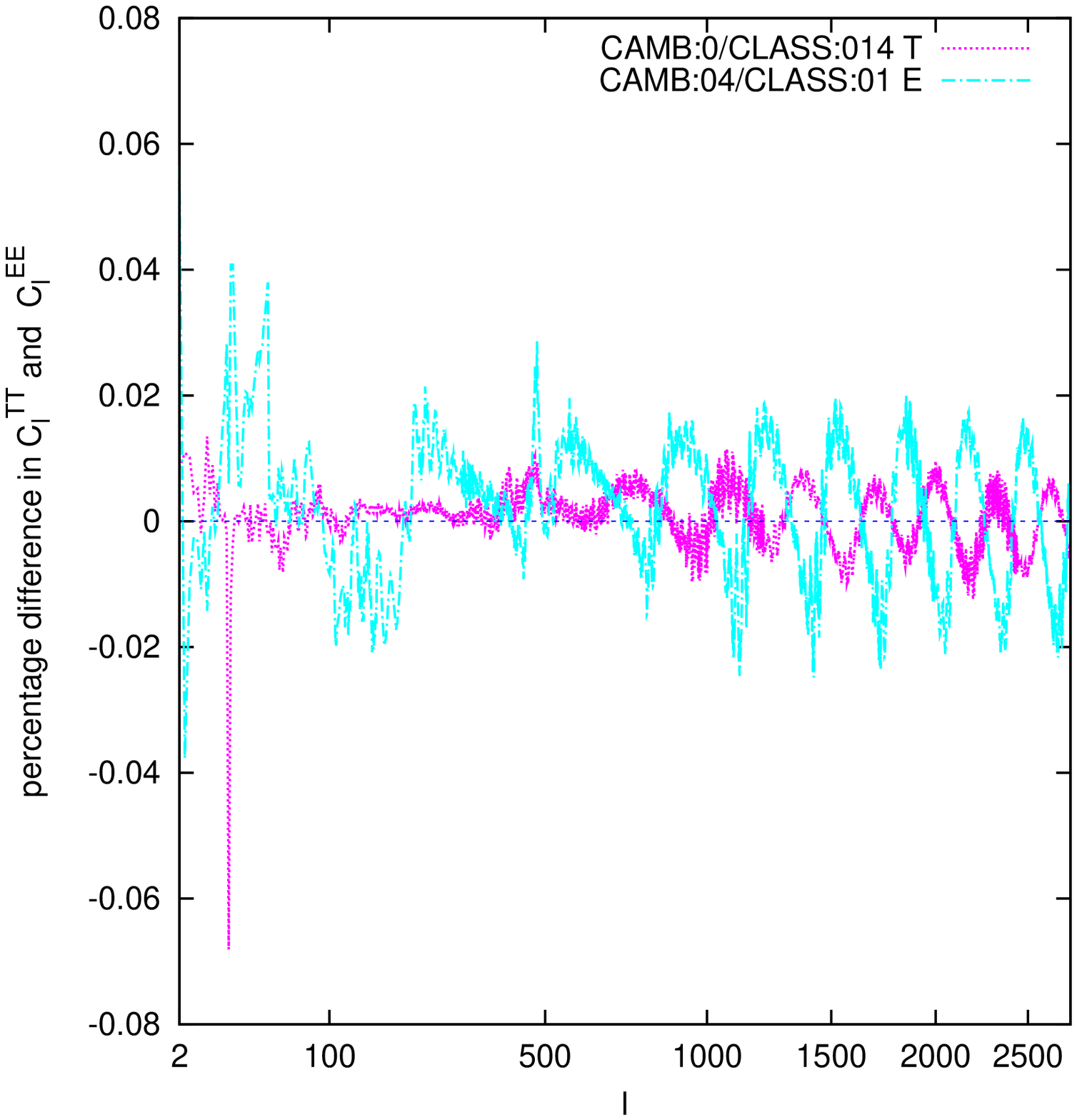}
\caption{\label{fig_cl} (Left) percentage difference between \CAMB and
\CLASS unlensed temperature anisotropy spectra $C_l^{TT}$, for
different precision settings for {\tt CAMB}, explained in the text. (Right)
best agreement reached in this work between \CAMB and {\tt CLASS}, for
unlensed temperature and E-polarisation $C_l$'s.}
}

The agreement between the two codes could probably be pushed even
further.  We strongly suspect that the dominant source of error
remains in the sector of thermodynamical quantities, especially for
{\tt CAMB}. Even {\it without changing any physical assumption about
  recombination,} the numerical accuracy could be increased by dealing
with issues of sampling, discontinuities, derivation algorithms,
etc. We did not pursue along this direction, which is somewhat
involved and probably not justified, since $0.01\%$ or $0.02\%$ errors
in the $C_l$'s are far below the sensitivity of realistic experiments
(we will see in section \ref{precision} that for Planck's effective
chi square, $\chi^2 \equiv -2\ln {\cal L}$, such errors lead to $\Delta
\chi^2 \sim 0.03$ only).

We conclude that current Boltzmann codes are able to predict the
$C_l$'s with an accuracy of the order of 0.01\% for a fixed
recombination history. As mentioned in the introduction, {\it physical
uncertainties on recombination}, on which Boltzmann codes have no
control, are clearly the major source of error in the computation of
CMB spectra.  

Of course, in order to reach such an accuracy, \CAMB and \CLASS are
both extremely slow (each run would require one or two days for each
model on a single processor; of course, we performed these runs in
parallel on many cores). The goal of section \ref{precision} will be
to degrade the precision and to speed up the two codes, with a
possibility to control the error by comparing with the reference
spectra derived in this section.

\subsection{Lensing potential and lensed CMB\label{lensed}}

The lensing potential multipoles $C_l^{\phi \phi}$ converge very
slowly as a function of the large-wavenumber cut-off in the
integration of transfer functions, parametrised in both codes by a
parameter $(k \tau_0)_\text{max}$ ($\tau_0$ being the conformal age of the
universe).  In order to obtain stable results at the 10\% level for
$C_l^{\phi \phi}$ around $l \sim 3000$, one would need to set $(k
\tau_0)_\text{max}$ to at least 15000.  Reaching such an accuracy at
$l=3000$ does not make much sense physically, since for such angular
scales the non-linear corrections are expected to be of the order of a
factor two. For comparison purposes, we fixed $(k \tau)_\text{max}$ to
12000 in both \CLASS (settings {\it [CAMB:04]}) and {\tt CAMB}. With
this setting, the absolute value of linear $C_l^{\phi \phi}$'s around
$l=3000$ is not per-cent accurate, but the results from the two codes
should in principle agree very well since the cut-off is the same in
the two cases. The error made by both codes on high-$l$ $C_l^{\phi
\phi}$'s is not worrysome, since it does not propagate to the lensed
CMB spectra.  This is easy to check explicitly: by varying $(k
\tau_0)_\text{max}$, we established that the setting $(k
\tau_0)_\text{max}=12000$ is indeed sufficient for obtaining stable
lensed CMB spectra in the range $l=2-3000$ at the 0.01\% level.

In view of computing the lensed $C_l$'s up to $l=3000$, we need the
$C_l^{\phi \phi}$'s up to a higher $l$'s, even if they do not need to
be very accurate. In the settings {\it [CAMB:05]}, we increase {\tt
  l\_max\_scalar} from 3000 to 4000. The settings {\it [CLASS:01]}
don't need to be changed, because \CLASS automatically increases
$l_\text{max}$ by an amount {\tt delta\_l\_max} when lensed $C_l$'s are
requested.  In the {\it [CLASS:01]} settings, {\tt delta\_l\_max} is
fixed to 1000, so that $l_\text{max}$ is also increased to 4000.

\CLASS computes the lensed CMB spectra from all-sky correlation 
functions~\cite{Challinor:2005jy}, i.e with the same method as {\tt CAMB},
but with a different numerical implementation written by S. Prunet,
based on quadrature weigths.

  We compare the $C_l^{\phi \phi}$ spectra from the {\it [CAMB:05]}
  and {\it [CLASS:01]} in figure \ref{fig_cll} (Left).  The two
  spectra agree at the level of $0.025\%$ in the range $3-4000$ (and
  $0.07\%$ for $l=2$): this is (by far) good enough in order to get
  accurate lensed temperature/polarisation multipoles. Indeed,
  sticking to the same precision settings, we compare the lensed
  temperature and polarisation spectra in the range $2-3000$ in figure
  \ref{fig_cll} (Right).  Nicely, the agreement remains as good as for
  the unlensed spectra (by $0.01\%$ for temperature excepted at
  $l=18$, and by $0.02\%$ for polarisation for $l\geq 40$).

\FIGURE{
\includegraphics[width=7cm]{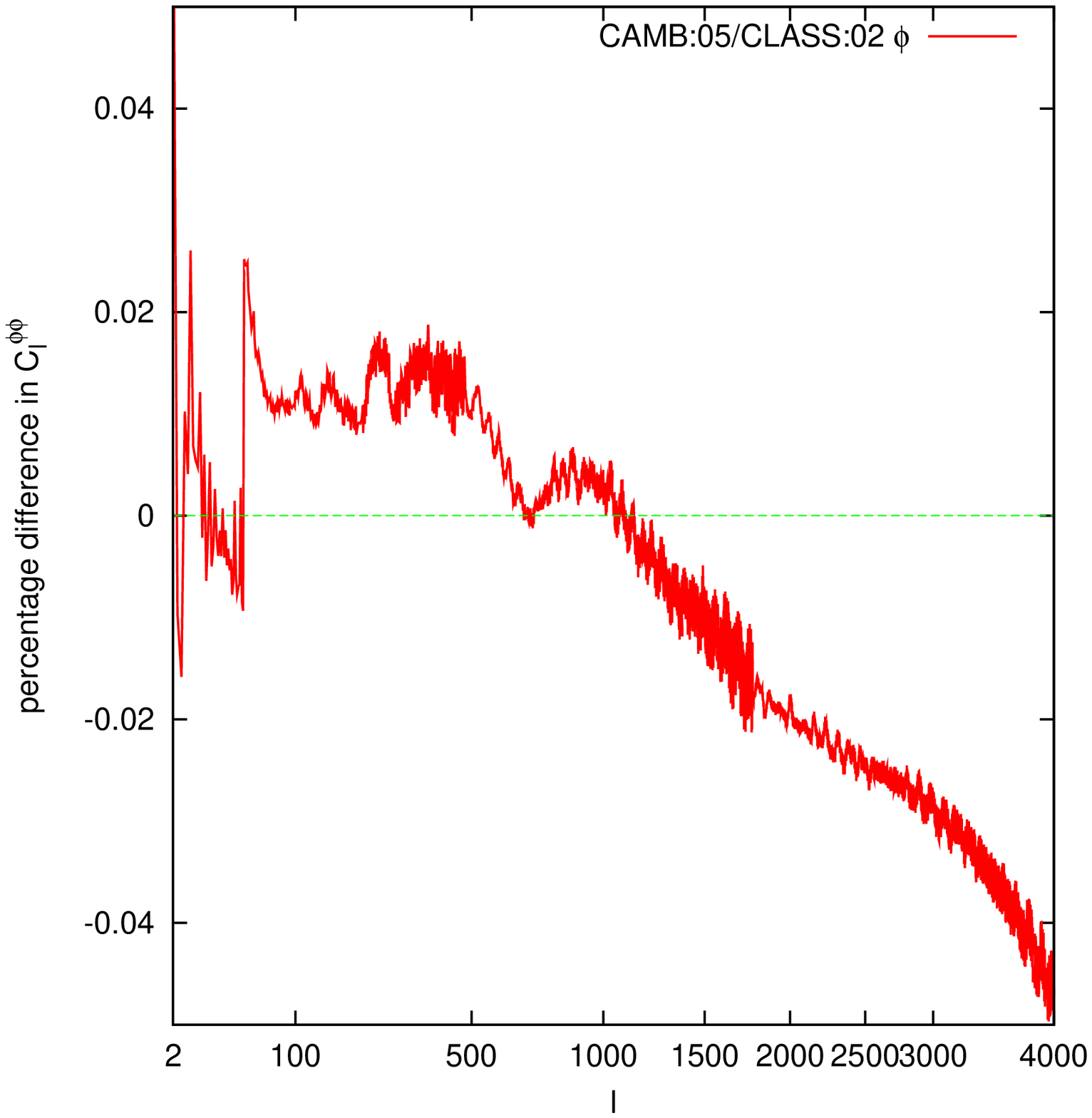}
\includegraphics[width=7cm]{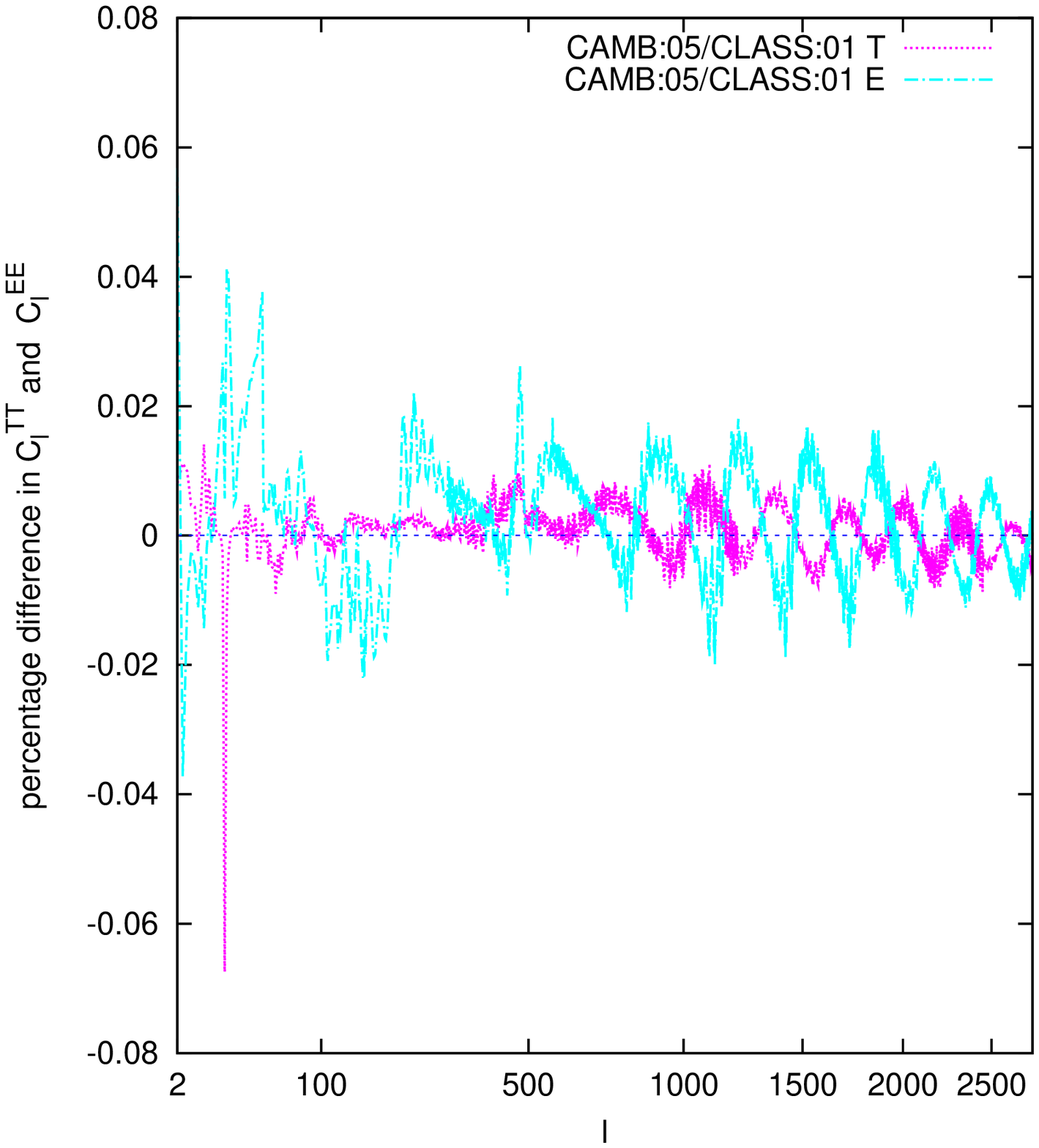}
\caption{\label{fig_cll} (Left) percentage difference between \CAMB and
\CLASS lensing potential anisotropy spectra $C_l^{\phi \phi}$, with the
precision settings explained in the text. (Right) percentage
difference between \CAMB and \CLASS lensed temperature anisotropy
spectra (temperature TT and polarisation EE).}
}

\subsection{Matter power spectrum\label{pk}}

In the standard cosmological scenario, the matter power spectrum can
be accurately described by linear theory up to about $k\sim
0.1\,h$Mpc$^{-1}$. Hence, one could argue that precise computations of
the matter power spectrum $P(k)$ for larger wavenumbers are
useless. Nevertheless, we want to push the comparison up to much
higher values, because the small-scale linear power spectrum is often
used as an input for different methods estimating the non-linear
spectrum: fitting formulas like HALOFIT \cite{Smith:2002dz}, codes
generating initial conditions for N-body simulations, algorithms
computing renormalised perturbations, etc. In this paper we focus on
the matter power spectrum in the range $10^{-4} h\,$Mpc$^{-1} <k <50
\, h\,$Mpc$^{-1}$.

The settings described in the previous subsections are sufficient in
order to get accurate predictions in the range from $10^{-4}$ to $1
h\,$Mpc$^{-1}$. Beyond that, it is necessary to increase the number of
Legendre multipoles for massless neutrino perturbations, $l_{\nu
\text{max}}$, in order to follow accurately neutrino free-streaming
during radiation domination (when their backreaction on metric
perturbations cannot be neglected). In {\tt CLASS}, $l_{\nu \text{max}}$
can be kept not too large thanks to the Ultra-relativistic Fluid
Approximation (UFA) described in ref.~\cite{class_approx}. The
precision parameter file {\tt pk\_ref.pre} released with \CLASS has
$l_{\nu \text{max}}$=150. Together with other settings, this is found to
be sufficient for $P(k)$ to converge at the $10^{-5}$ level, up to at
least $k_\text{max}=50 \, h\,$Mpc$^{-1}$. We refer to \CLASS runs with
the accuracy file {\tt pk\_ref.pre} as {\it [CLASS:02]}.

We compared the resulting $P(k)$ with that derived from \CAMB with the
following settings, that we call {\it [CAMB:06]}:

\vspace{0.2cm}

\begin{tabular}{rcl}
{\tt transfer\_high\_precision} &=& T\\
{\tt accuracy\_boost}          &=& 3\\
{\tt l\_accuracy\_boost}        &=& 3\\
{\tt do\_late\_rad\_truncation}          &=& T
\end{tabular}

\vspace{0.2cm}

\noindent Other relevant parameters have been set to the same values
as in the previous {\it [CAMB:05]} settings (including {\tt
  nthermo}=100'000).  The result of the comparison, shown in
fig.~\ref{fig_pk} (Left), exhibits a few features at the level of
0.1\%, plus a very sharp raise above $1 h\,$Mpc$^{-1}$.  We tried to
turn off {\tt CAMB}'s {\tt do\_late\_rad\_truncation}
approximation\footnote{this approximaxion introduces a small error in
  the matter power spectrum, mainly due to the fact that it neglects
  the impact of reionisation on the baryon perturbations. The
  equivalent approximation in \CLASS (the Radiation Streaming
  Approximation described in ref.~\cite{class_approx}) takes into
  account this effect and is more model-independent.} and to increase
accuracy boost parameters until the matter power spectrum converges up
to the $10^{-5}$ level. This occurs for the settings

\vspace{0.2cm}

\begin{tabular}{rcl}
{\tt transfer\_high\_precision} &=& T\\
{\tt accuracy\_boost}          &=& 5\\
{\tt l\_accuracy\_boost}        &=& 8\\
{\tt do\_late\_rad\_tr-unction}          &=& F
\end{tabular}

\vspace{0.2cm}

\noindent that we call {\it [CAMB:07]}. The large value of the {\tt
  l\_accuracy\_boost} parameter reflects the need to increase $l_{\nu
  \text{max}}$, especially in absence of an UFA approximation. At this
  point, the two matter power spectra agree up to the 0.01\% level up
  to $1 h\,$Mpc$^{-1}$, but the sharp raise persists in the large-$k$
  limit. A careful inspection shows that this effect comes entirely
  from the fact that \CAMB neglects the impact of reionisation on the
  baryon sound speed $c_s^2(z)$, while \CLASS sticks to the equations
  from \cite{Ma:1995ey} which assume thermal equilibrium between
  baryons and electrons.  Indeed, \CAMB infers the baryon sound speed
  from the free electron fraction immediately after calling
  \RECFAST{}, while \CLASS uses the free electron fraction already
  corrected by the effect of reionisation.  Hence, at low redshift,
  $c_s^2$ is one order of magnitude larger in {\tt CLASS}; the Jeans
  instability of baryons is then responsible for a cut-off in the
  matter power spectrum not far from $k\sim10 h\,$Mpc$^{-1}$. We
  checked in fig.~\ref{fig_pk} (Right) that if we run \CLASS with the
  same cosmological parameters but reionisation switched off, the
  results of {\it [CAMB:07]} and {\it [CLASS:02]} agree at the 0.01\%
  level at least till $k= 50 h\,$Mpc$^{-1}$.
  
Of course, in the non-linear power spectrum and near the wavenumber
$k\sim50 h\,$Mpc$^{-1}$, the cut-off is erased by the transfer of
large-scale to small-scale power. So, in practice, this difference
between the two codes (which both oversimplify the baryon evolution
during reionisation in different ways) is not a problem.

\FIGURE{
\includegraphics[width=7cm]{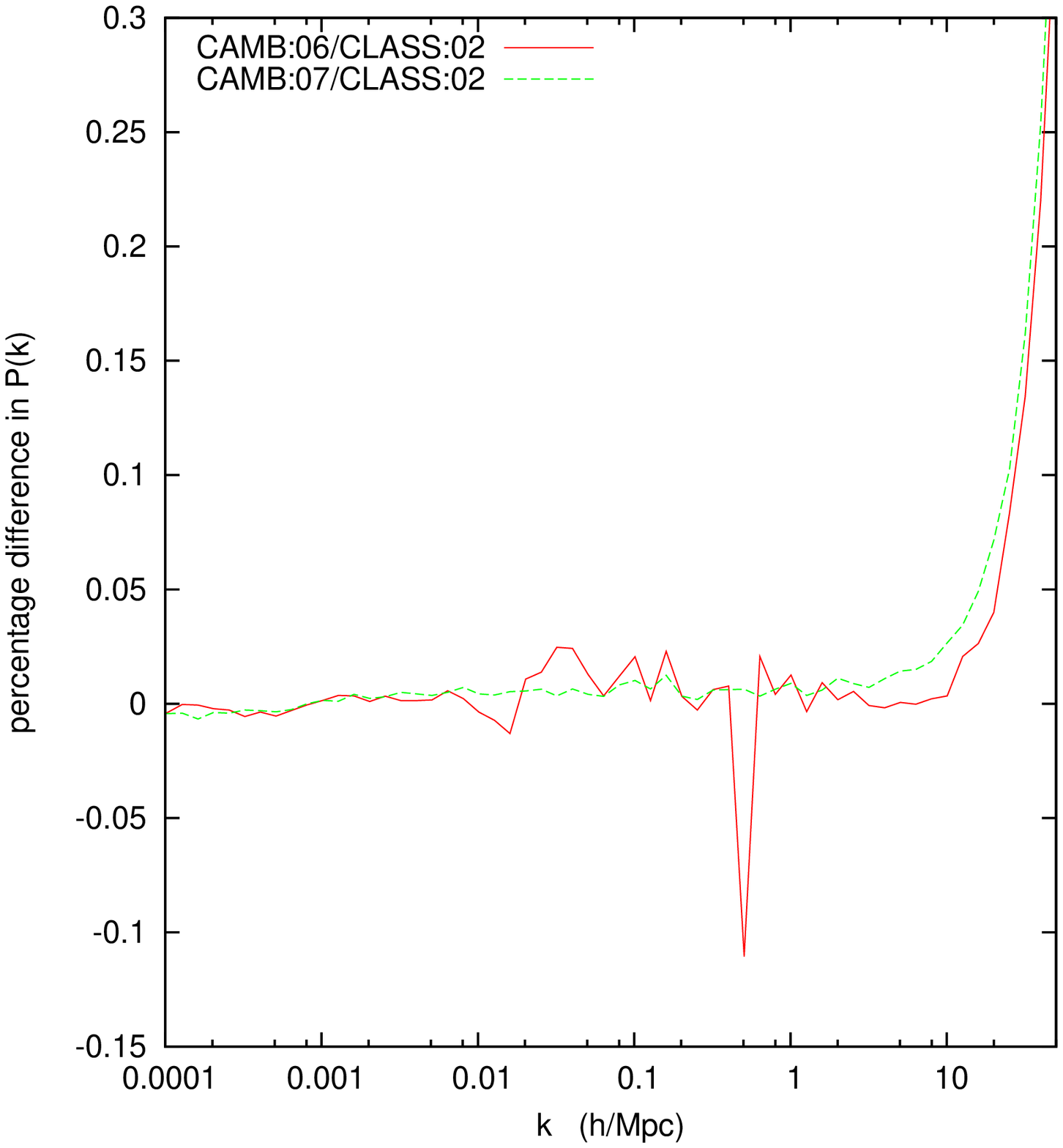}
\includegraphics[width=7cm]{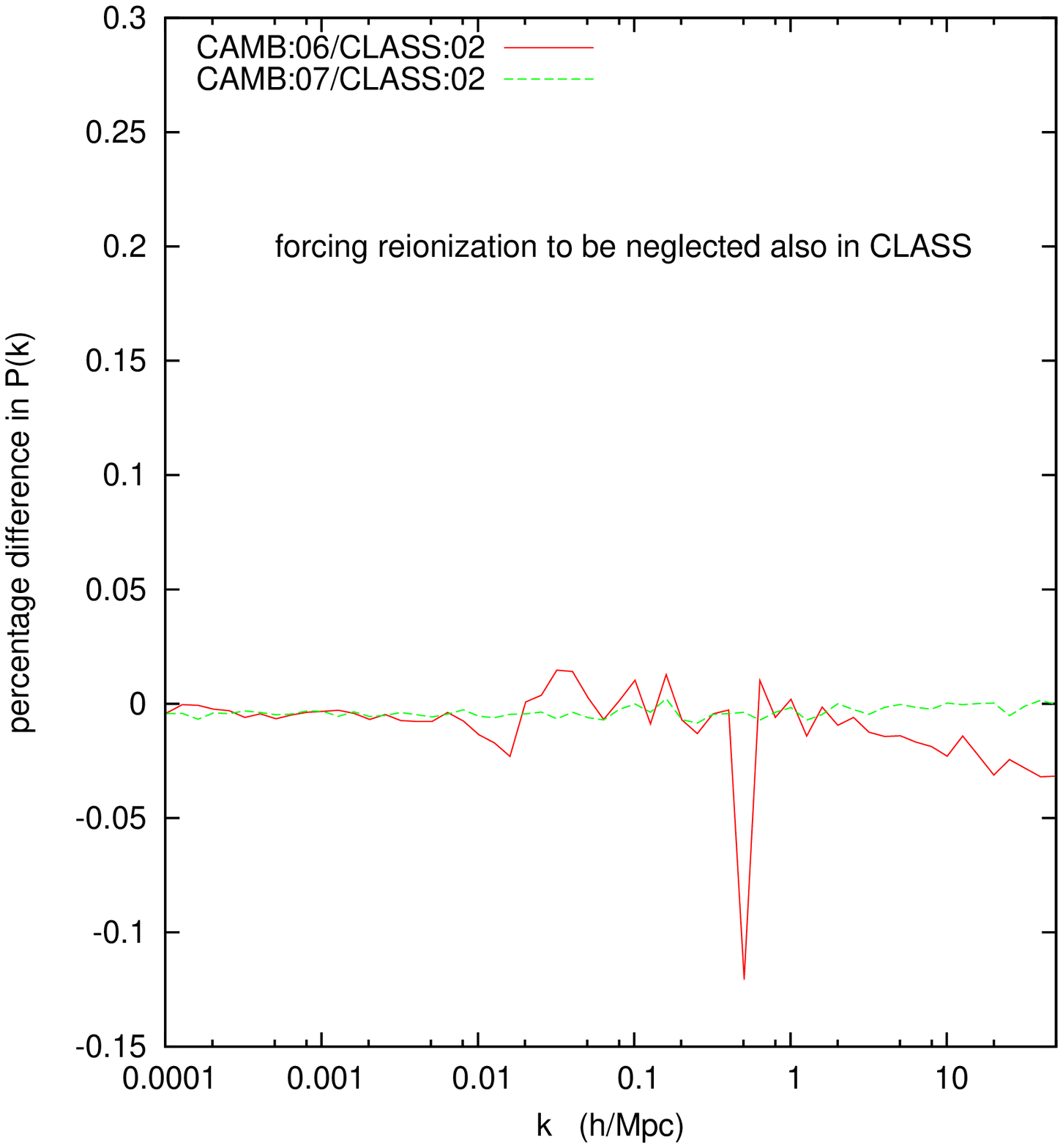}
\caption{\label{fig_pk} (Left) percentage difference between \CAMB and
\CLASS matter power spectrum $P(k)$, with the precision settings
explained in the text. (Right) same if we force \CLASS to neglect the
impact of reionisation on the baryon sound speed, like \CAMB does.}
}

\section{Precision settings\label{precision}}

The settings {\it [CLASS:01]} and {\it [CLASS:02]} can be used to
produce reference spectra, and then to calibrate different degraded
precision settings, adapted to the user's goal. In the distribution of
\CLASS v1.0, these settings correspond to the precision files {\tt
cl\_ref.pre} and {\tt pk\_ref.pre}. We recall that the first one is
sufficient for getting 0.01\% accurate $C_l^{TT}$'s and $P(k)$'s in
the range $2\leq l\leq3000$ and $k\leq 1 h\,$Mpc$^{-1}$ (for
$C_l^{EE}$ the error is twice larger). The second settings is only
necessary for maintaining such an accuracy on the linear $P(k)$ till
$k\leq 50 h\,$Mpc$^{-1}$.

Under many circumstances, a Boltzmann code user wants to be sure that
the error made on the $C_l$'s is smaller than some level (e.g., than
the level of the new physical effect that he/she wants to study). For
this purpose, we derived three settings achieving respectively 0.1\%,
0.2\% or 0.3\% accuracy on the $C_l$'s up to $l=3000$: they are
contained in the precision files {\tt cl\_permille.pre}, {\tt
  cl\_2permille.pre}, {\tt cl\_3permille.pre} distributed together
with the code. For each of these settings, the precision on the matter
power spectrum for $k\leq 1 h\,$Mpc$^{-1}$ is even better than that on
the $C_l$'s.

However, for parameter extraction and large Monte-Carlo runs, these
settings are not optimal: they do not minimise the computing time for
a given data sensitivity. In principle, for each new data set, one
could derive optimal precision settings for {\tt CLASS}. The way to proceed
is to compute some reference spectra using e.g. the {\tt cl\_ref.pre}
file, and then to degrade the precision while keeping the likelihood
of the output spectra below a given level (given instrumental noise
and taking the fiducial spectra to be the reference ones).

Here, we illustrate this approach by using a simplified Planck
likelihood. In the future, it will be easy to derive
e.g. ``Planck+SDSS'' or ``Core+Euclid'' precision files. This task is
made easier if one assumes that each accuracy parameter degrades the
precision independently of the others, i.e. that no complicated
combinations of the accuracy parameters can make the code faster for a
fixed precision. Under this very plausible assumption, deriving
precision settings is trivial, even for somebody who is not expert in
the code: one should consider the accuracy parameters {\it one after
each other}, loop over several values of these parameters, compute the
$\Delta \chi^2$ relative to the reference spectrum, and take the
largest possible value of each accuracy parameter compatible with a
given order of magnitude for $\Delta \chi^2$ (here $\chi^2$ is defined
to be $-2 \ln {\cal L}$, and ${\cal L}$ is the likelihood of a
combination of experiments). The role of each precision parameter is
summarised in the comments of the \CLASS file ${\tt
include/common.h}$, where all precision parameters are declared within
a structure called {\tt precision}.

For illustration, we assume here a simplistic Planck likelihood based
on three channels, with a gaussian isotropic noise and a fraction of
the sky $f_\text{sky}=0.8$. Details on this likelihood can be found e.g. in
\cite{Perotto:2006rj}. The instrumental noise from Planck increases
exponentially above $l=2500$, but throughout this comparison exercise we
compute the $\chi^2$ up to 3000, even if the last mutipoles do not
contribute much. In order to obtain some $C_l$'s at $l=3000$, we must
keep ${\tt l\_max\_scalar}$ higher than 3250 in \CAMB{}.

We first checked that the \CLASS and \CAMB
reference CMB spectra presented in the previous section differ by
$\Delta \chi^2=0.027$: we conclude that given the current accuracy of
Boltzmann codes, the effective $\chi^2$ obtained with Planck will only
be accurate up to 0.03. Fortunately, this precision is by far sufficient for
extracting cosmological parameters in a robust way, without
introducing any sizeable ``Boltzmann code bias''.

We then degrade the precision of the two codes as much as possible in
order to keep the error below $\Delta \chi^2 \sim 0.1$, 1 or 10.
During this step, we compute the $\Delta
\chi^2$ relative to the \CLASS reference spectra, but it would be
equivalent to define it with respect to the \CAMB one. Let us stress
that we are not claiming here that one of the above settings should be
effectively used when fitting Planck data. The only way to check which
precision setting is sufficient is to perform a full parameter
extraction from the data with different settings, and keep the
loosest/fastest settings such that no bias on the observed parameters
is observed. A difference $\Delta \chi^2 \sim 1$ sounds very large at
first sight, but it is probably small enough for an accurate parameter
extraction. This issue will be studied more in details in future work.
Here, we only regard the $\Delta \chi^2$'s as a measure
of accuracy. Note that $\Delta \chi^2 \sim 1$ roughly corresponds to
0.1-0.2\% accuracy on the $C_l$'s, except on very large and very small
angular scales.

The resulting settings for \CLASS are stored in precision files {\tt
chi2pl0.1.pre}, {\tt chi2pl1.pre}, {\tt chi2pl10.pre}, distributed
together with the code.  For {\tt CAMB}, we found the optimal settings
summarised in Table
\ref{camb-set}.  Note that for a fair comparison, we kept the
parameter ${\tt nthermo}$ of {\tt modules.f90} fixed to $100'000$ (see
section \ref{unlensed}) at least in the highest accuracy settings:
this lowers the $\Delta \chi^2$ by about 0.02 without affecting the
velocity of the code. For $\Delta \chi^2 \geq 1$, it is sufficient to keep
{\tt CAMB}'s default values of the Bessel function sampling, but
for $\Delta \chi^2=0.1$ we were forced to open the file {\tt
bessels.f90} and to lower $\Delta x$ for $x>25$ down to 0.8. Note that
when we keep all default accuracy settings in \CAMB{} (summarised in the last
column of Table \ref{camb-set}) we find $\Delta
\chi^2=368$ (the increase from $\Delta \chi^2=10$ to 368 
is only caused by lowering {\tt accuracy\_boost} from 1.5 to 1, 
and {\tt l\_sample\_boost} from 1.2 to 1).

\TABLE{
\begin{tabular}{l|c|c|c|c|c}
 & reference & $\Delta \chi^2 \sim 0.1$ & $\Delta \chi^2 \sim 1$ &
 $\Delta \chi^2 \sim 10$ & default\\
\hline
{\tt l\_max\_scalar}           & 4000 & 3300 & 3300 & 3300 & 3300\\
{\tt k\_eta\_max\_scalar}      & 12000 & 6600 & 4500 & 3500 & 3500\\
{\tt accurate\_BB}             & T & F & F & F & F\\
{\tt accurate\_polarization}   & T & F & F & F & F\\
{\tt accurate\_reionization}   & T & T & T & F & F\\
{\tt do\_late\_rad\_truncation}& F & T & T & T & T\\
{\tt accuracy\_boost}          & 12.0 & 3.0 & 2.0 & 1.5 & 1.0\\
{\tt l\_accuracy\_boost}       & 4.0 & 2.0 & 1.5 & 1.2 & 1.0\\
{\tt l\_sample\_boost}         & 3.0 & 1.5 & 1.0 & 1.0 & 1.0\\
{\tt nthermo} in {\tt modules.f90} & 100'000 & 100'000 & 100'000 & 10'000 & 10'000\\
$\Delta x$ for $x>25$ in {\tt bessels.f90} & 0.2 & 0.8 & 1.0 & 1.0 & 1.0\\
\end{tabular}
\caption{\label{camb-set} \CAMB accuracy setting in the reference model,
and for levels of precisions explained in the text.}
}

\section{Compared performances\label{performance}}

\FIGURE{
\includegraphics[width=10cm]{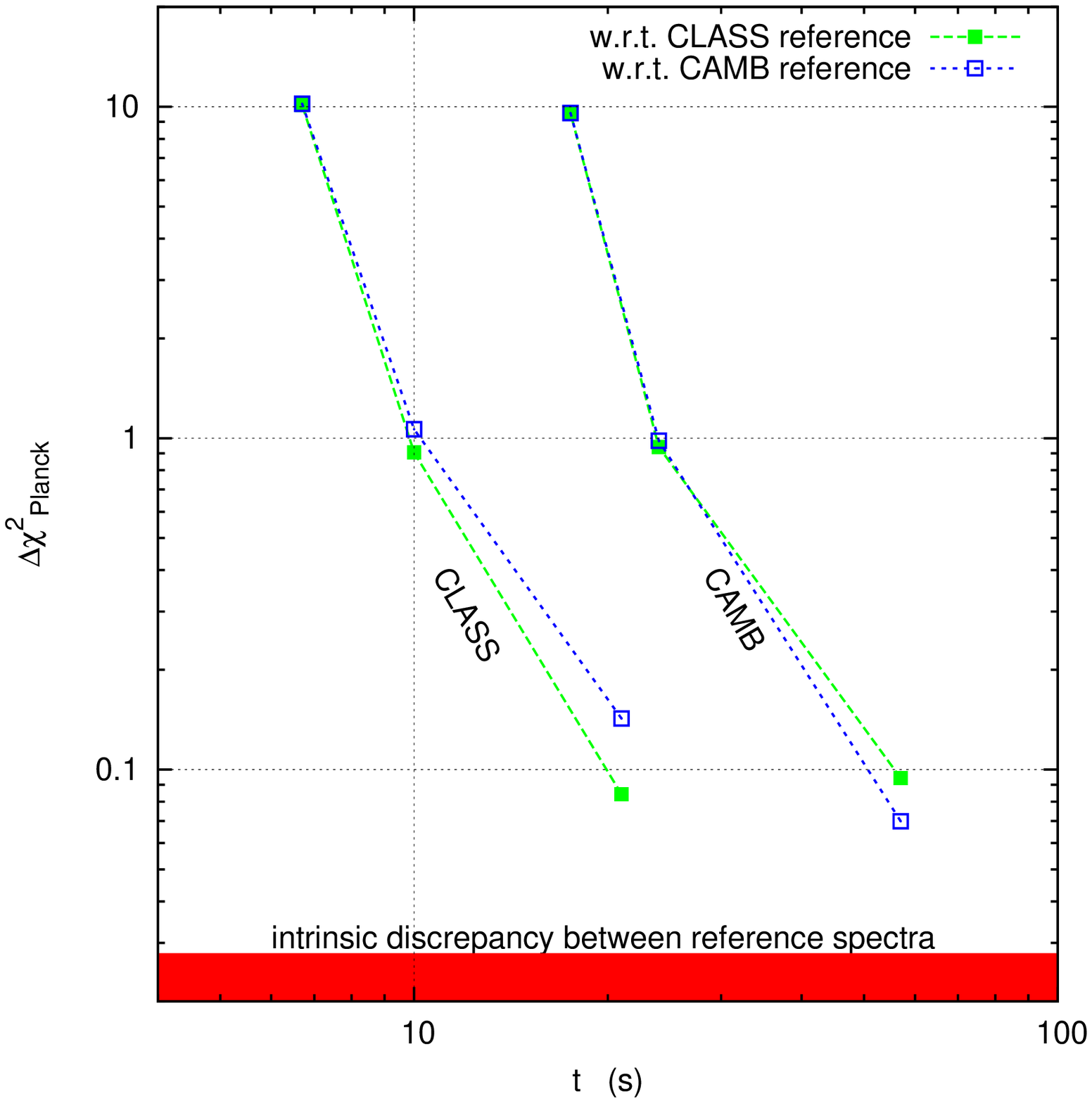}
\caption{\label{perf} Precision versus running time for \CAMB and
{\tt CLASS}, for different accuracy setting targeted at $\Delta
\chi^2 \sim 0.1$, 1 or 10, where $\chi^2 = -2 \ln
{\cal L}$ is the effective $\chi^2$ of a simplistic Planck likelihood
${\cal L}$ (including temperature and polarisation). This $\Delta
\chi^2$ is evaluated by comparing either with the \CLASS reference
spectra (solid squares) or the \CAMB ones (empty squares). The \CAMB
curves are roughly shifted by a factor 2.5 along the time axis with
respect to the \CLASS ones.}
}

Having calibrated the precision of \CLASS and {\tt CAMB}, we can finally
compare their preformances in terms of execution time. The result may
depend slightly on several hardware and software issues. Of course, we
always performed our runs on the same computer. Both codes are
parallelised (\CLASS has many parallel zones in the {\tt perturbation},
{\tt bessel}, {\tt transfer}, {\tt spectra} and {\tt lensing} modules), but the
comparison is more robust if we limit the execution of both codes to
a single CPU. The performances could be affected by
different choices of compilers and optimisation flags. For both codes
we use a gnu compiler ({\tt gfortran} for {\tt CAMB}, {\tt gcc} for {\tt CLASS})
with the flag {\tt -O4}.  Different choices are unlikely to change our
conclusions by a significant amount.

In figure \ref{perf} and for the precision settings found in the
previous section, we plot the code's precision (quantified by $\Delta
\chi^2$) as a function of its execution time. The horizontal axis is
the running time in seconds on our processor; using a different
computer, this axis would just be renormalised by some number. The
intrinsic limitation of the codes is given by the line $\Delta
\chi^2=0.027$ (since this corresponds to the discrepancy between the
reference spectra from \CLASS and {\tt CAMB}). For each of the codes, the
squares mark $(t, \Delta \chi^2)$ for each of the three optimal
precision settings found in the previous section; filled squares
correspond to a $\Delta \chi^2$ computed with respect to the \CLASS
reference spectra, and empty squares to that with respect to the \CAMB
reference spectra. For a given $\Delta \chi^2$, \CLASS is always $\sim
2.5$ faster.

We did not push the exercise further. We could have included a matter
power spectrum likelihood. We believe that the results would not
change considerably, because computing an accurate $P(k)$ up to the
non-linearity scale ($k \sim 0.1$ or 0.2 $h$Mpc$^{-1}$) requires a
marginal increase in computation time with respect to computing the
lensed CMB (at least, this is true for $\Lambda$CDM models). Would
there be a significant change in the speed ratio, it could only be in
favor of {\tt CLASS}, because the approximation scheme RSA discussed in
\cite{class_approx} allows to speed up considerably the integration
during matter/$\Lambda$ domination without affecting the precision of
the $P(k)$, while the comparable approximation in \CAMB (controlled by
{\tt do\_late\_rad\_truncation}) affects slightly the baryon evolution
during reionisation \cite{class_approx} and alters the final $P(k)$
(in section \ref{pk}, we were forced to keep this approximation turned
off for getting accurate $P(k)$'s).

For extended cosmological scenarios (including e.g. tensors, spatial
curvature, etc.), the comparison of \CLASS and \CAMB performances will
be presented in future communications. For models with massive
neutrinos or other non-cold dark matter relics, \CLASS includes some
very fast approximation schemes which performances are discussed in
\cite{class_ncdm}.

\section{Conclusions}

The main result of this paper is the fact that independent Boltzmann
codes agree at a level of precision which had not been investigated
before. This definitely proves that current Boltzmann code are by far
accurate enough for analysing Planck data, as well as the following
generation of CMB and Large Scale Structure experiments.  In future
cosmological parameter extractions, a small level of theoretical
errors will remain, coming from uncertainties on the recombination
history, on the details of reionisation, on our ability to model
foregrounds, etc. Instead, the error arising from Boltzmann codes is
now proved to be fully under control.

The main objective of \CLASS is to offer an accurate, friendly and
flexible environment for coding extended cosmological models and
comparing them with cosmological data. The issue of speed is less
important, but the observation that \CLASS is faster than the best
existing alternatives (by a factor $\sim$2.5 for minimal $\Lambda$CDM) is
certainly positive. It is not the purpose of this paper to discuss the
reasons for which \CLASS is slightly faster than other codes. Since
they were written independently, \CLASS and \CAMB differ by hundreds
of details. The most important ones (approximation schemes,
integrators, step sizes, etc.) are discussed in companion papers
\cite{class_general,class_approx}. We believe that they all slightly
contribute to the performances of the code.

\section*{Acknowledgments}

We would like to thank A.~Lewis for very useful comments on this
manuscript.  In Refs. \cite{class_general, class_approx}, we already
thanked several colleagues who stimulated the \CLASS project with
various forms of input. The work of S.~Prunet in the lensing
module was particularly important for the achievements presented in
section \ref{lensed} of this work, and the ndf15 integrator developed
specially for \CLASS by T.~Tram is one of the ingredients leading to
the good performances of the code. Running with high-precision
settings is only practical on machines with many cores and large
memory: we wish to thank M. Shaposhnikov for providing us with a brand
new 48-core PC at EPFL, and the Institut d'Astrophysique de Paris for
giving us access to the {\tt horizon9} machine.

More generally, we wish to express our admiration for the authors of
previous Boltzmann codes, who were able to develop fast and compact
packages so many years before the present attempt - the fact of
arriving later, with excellent implementations already available on
the market, made everything easier.

%%%%%%%%%%%%%%%%%%%%%%%%%%%

\end{document}